\documentclass[twocolumn,superscriptaddress,longbibliography,aps,prl,preprintnumbers]{revtex4-2}

\usepackage[normalem]{ulem}
\usepackage{graphicx}
\usepackage{bm}
\usepackage{color}
\usepackage{epsfig}
\usepackage{amsmath}
\usepackage{amssymb}
\usepackage{wrapfig}

\usepackage{hyperref}
\hypersetup{
    colorlinks=true,      
    urlcolor=blue,
    citecolor=blue,
    linkcolor=blue
}

\begin{document}

	\title{Probing Fermi surface topology by ultrafast pump pulse dynamics}
	
	\author{Debamalya Dutta}
	\author{Kush Saha}
	\affiliation{National Institute of Science Education and Research, Jatni, Odisha 752050, India\\Homi Bhabha National Institute, Training School Complex, Anushakti Nagar, Mumbai 400094, India}
	
	
	\begin{abstract}
		
		We present a dynamical approach to detect changes in Fermi surface topology in a two-band model. Specifically, we show that the system's response to a low intensity light pulse can precisely identify topological Lifshitz transitions. At a suitable frequency, the light resonantly couples valence and conduction electrons, leading to an oscillation in the interband coherence term. This, in turn, generate a persistent oscillatory current which survives even after the end of the pulse.
		Notably, the relative amplitude of the oscillatory current during the pulse with that of the post-pulse reaches a minimum when the Fermi energy aligns with the saddle point, providing a robust framework for dynamically identifying Lifshitz transitions.
		
	\end{abstract}
	
	\maketitle
	
	{\em Introduction.-} 
	A plethora of unconventional electronic behaviors and anomalies in materials have been linked to changes in Fermi surface (FS) topology across the saddle point of the band spectrum, a phenomenon known as the Lifshitz transition (LT)~\cite{10.1063/1.4974185,Volovik_2018,BLANTER1994159,lifshitz1960anomalies}. In $WTe_2$~\cite{10.1063/5.0070914}, finite temperature resistivity is found to show anomalous peak at a specific temperature when the FS is near the saddle point.  Similarly, pressure-induced FS variations in $SnSe$~\cite{PhysRevLett.122.226601} 
	enhance thermoelectric power, as a manifestation of the LT. Additionally, LT has been found to coincide with the onset of superconductivity~\cite{PhysRevB.78.205431} in electron-doped iron arsenic superconductors~\cite{Liu2010}. Various experimental techniques such as field modulation technique with the use of superconducting magnet, cyclotron resonance (CR)~\cite{PhysRevLett.108.017602} absorption experiment, transverse electron focusing experiments \cite{PhysRevLett.133.096601} are proven to be successful direct or indirect techniques to understand the geometry of the FS.
	
	While various equilibrium transport properties such as conductivity, quantum oscillations (de Haas–van Alphen effect ~\cite{10.21468/SciPostPhysProc.11.019,PhysRevB.107.075120,PhysRevB.104.085412}, Shubnikov–de Haas effect ~\cite{B_L_Zhou_1981,Nowakowska2023,PhysRevB.108.235155,10.1063/1.5049160}) and recently developed Andreev state transport~\cite{PhysRevLett.130.096301} are known to provide insights into the Fermi surface topology and Lifshitz transition, studies on the nonequilibrium dynamical phenomena linked to the Fermi surface remain limited. A recent experimental study demonstrated that an ultrashort laser pulse can be used as a dynamical route to control Fermi surface topology in ultra-fast time scales\cite{doi:10.1126/sciadv.abd9275}. However, diagnostic probe or systematic approach to identify Lifshitz transition in out-of-equilibrium domain is yet to be addressed. In particular, we ask if the response of a weak light pulse can probe the Lifshitz transition in a quantum material. 
	
	In this work, we demonstrate a dynamical approach to probe FS in a two-band model using the current response to an external light pulse. We provide an approximate analytical expression for interband coherence term which is found to be oscillatory under resonant coupling between the conduction and valence electrons. We find that this feature is then reflected in the current during the pulse and continues even after the end of the pulse with reduced but constant amplitude. Remarkably, we find that the ratio between the maximum amplitude of the oscillation during the pulse and the amplitude of the post-pulse oscillation can encode the signature of topological Lifshitz transition.

	{\em Methodology--} 
	We consider a two-band model Hamiltonian exposed to a low intensity light field $\vec{E}(t)=-\partial_t\,\vec A(t)$. The field couples to the momentum through the vector potential, $\vec A(t)$ in the velocity gauge as $\mathbf{k}\to \mathbf{k}-\frac{q\vec A(t)}{\hbar}$. The response to such an externally time-dependent field can be obtained using the total current
	\begin{equation}
	\mathfrak{J}(t)=\int \mathcal{J}(\mathbf{k},t)\,\frac{d\mathbf{k}}{(2\pi)^2}\,\label{eq:total_current},
	\end{equation}
	where $
	\mathcal{J}(\mathbf{k},t)=\frac{q}{\hbar}\mathrm{Tr}(\hat{\rho}(\mathbf{k},t)\nabla_{\mathbf{k}}\hat{\mathcal{H}}(\mathbf{k},t))\,
	$ with $\hat{\mathcal{H}}(\mathbf{k},t)$ being the time-dependent Hamiltonian and  $\hat{\rho}$ is the density matrix. For a two-band model $\hat{\rho} (\mathbf{k},t)$ is written as
	\begin{align}
	\hat{\rho}(\mathbf{k},t)=\begin{pmatrix}
	\rho_{11}(\mathbf{k},t) & \rho_{12}(\mathbf{k},t)\\
	\rho_{21}(\mathbf{k},t) & \rho_{22}(\mathbf{k},t)
	\end{pmatrix}\,,
	\end{align}
	where $\rho_{mm}(\mathbf{k},t)$ refers to the population $n_m(\mathbf{k},t)$ of the $m$-th band, and $\rho_{m,m'}(\mathbf{k},t)$ is the interband coherence term, denoted by $\pi(\mathbf{k},t)$.
	
	To find $\hat{\rho}(\mathbf{k},t)$, we solve the Liouville–von-Neumann equation (LVNE)
	\begin{equation}
	i\hbar \dot{\hat{\rho}}({\mathbf{k}},t)=[\mathcal{\hat{H}}({\mathbf{k}},t),\hat{\rho}({\mathbf{k}},t)]\label{eq:LVNE}\,,
	\end{equation} 
	with the initial conditions as $n_g(\mathbf{k},t_0)=f(E_g(\mathbf{k}))$, $n_e(\mathbf{k},t_0)=f(E_e(\mathbf{k}))$ and $\pi(\mathbf{k},t_0)=0$, where $E_g(\mathbf{k})$ and $E_e(\mathbf{k})$ are the energies corresponding to valence and conduction bands, respectively and $f$ is the Fermi-Dirac distribution function.

	Equation~(\ref{eq:LVNE}) leads to first order differential equations for the populations of conduction and valence bands as
	\begin{align}
	&\hspace*{-0.4cm}\hbar\,\dot{n}_{g}(\mathbf{k},t)=iq\vec{E}(t).\left[\pi^*(\mathbf{k},t)\vec{d}_{e,g}(\mathbf{k})-\pi(\mathbf{k},t)\vec{d}_{g,e}(k)\right]\,,\label{eq:4ng}\\
	&\hspace*{-0.4cm}\hbar\,\dot{n}_{e}(\mathbf{k},t)=iq\vec{E}(t).\left[\pi(\mathbf{k},t)\vec{d}_{g,e}(\mathbf{k})-\pi^*(\mathbf{k},t)\vec{d}_{e,g}(\mathbf{k})\right]\,,
	\label{eq:4nc}
	\end{align}
	where $\vec{d}_{m,m'}(\mathbf{k})=i\int u_{m\,\mathbf{k}}^*(x)\partial_{\mathbf{k}} u_{m'\,\mathbf{k}}(x)\,dx$ is the dipole matrix element (see Supplementary Materials). For $m\ne m'$, we obtain interband coherence term
	\begin{equation}
	\hbar\,\dot{\pi}(\mathbf{k},t)=-i\mathcal{X}(\mathbf{k},t)\pi(\mathbf{k},t)+iq\vec{E}(t).\vec{d}_{e,g}(\mathbf{k}) n(\mathbf{k},t), \label{eq:4pi2}\,
	\end{equation}
	where 
	$\mathcal{X}(\mathbf{k},t)=\left[\epsilon(\mathbf{k})+q\vec{E}(t)\left\{d_{e,e}(\mathbf{k})-d_{g,g}(\mathbf{k})\right\}\right]\nonumber$ with $\epsilon(\mathbf{k})=E_e(\mathbf{k})-E_g(\mathbf{k})$,
	and $n(\mathbf{k},t)=n_e(\mathbf{k},t)-n_g(\mathbf{k},t)$ is the population difference. The Eqs.~(\ref{eq:4ng}), (\ref{eq:4nc}) and (\ref{eq:4pi2}) are known as the Semiconductor Bloch equations (SBE)s.
	
	{\em Solution--} We primarily focus on the solution of interband coherence term in Eq.~(\ref{eq:4pi2})  which turns out to be key quantity of the present study.  The solution of Eq.~(\ref{eq:4pi2}) can be written as
	\begin{equation}
	\pi(\mathbf{k},t)=\mathcal{I}_1(\mathbf{k},t)e^{-\frac{i}{\hbar}\int_{t_0}^t\mathcal{X}(\mathbf{k},T_1)dT_1}\label{eq:4pi3}\,,
	\end{equation}
	where
	\begin{align}
	\mathcal{I}_1(\mathbf{k},t)&=\int_{t_1\leq t_0}^t\mathcal{K}_1(\mathbf{k},t)e^{\frac{i}{\hbar}\int_{t_0}^{T_2}\mathcal{X}(\mathbf{k},T_1)dT_1}dT_2\label{eq:4int1}
	\end{align}
	with $\mathcal{K}_1(\mathbf{k},t)=\frac{i}{\hbar}q\vec{E}(t).\vec{d}_{e,g}(\mathbf{k},t)n(\mathbf{k},t)$.
	The term $\mathcal{X}$ is in general complex and becomes purely real at the end of the pulse. Hence $\pi({\bf k},t)$ is oscillatory even after the end of the pulse and the amplitude of the oscillation depends on $\mathcal{I}_1(\mathbf{k},t)$. This amplitude of {\it the post-pulse oscillation} can be used to identify saddle point in the spectrum as will be evident shortly. 
	
	To find an estimate of the amplitude of $\mathcal{I}_1$, we expand $\mathcal{K}_1$ in Fourier components as
	\begin{equation}
	\mathcal{K}_1(\mathbf{k},t)=\sum_{n}\widetilde{\mathcal{K}}_1(\mathbf{k},\omega_n)e^{-i\omega_n t}\,.
	\end{equation}
	For a weak field $\vec{E}(t)$, $\mathcal{K}_1(\mathbf{k},t)$ can be approximated as $\mathcal{K}_1(\mathbf{k},t)\simeq \widetilde{\mathcal{K}}_1(\mathbf{k},\omega_{\mathrm{pulse}})e^{-i\,\omega_{\mathrm{pulse}}\, t}$ as the system is mainly governed by the pulse frequency $\omega_{\mathrm{pulse}}$ of the applied field. 
	Subsequently, $\mathcal{X}(\mathbf{k},t)$ can be approximated as $\epsilon(\mathbf{k})$ at such low field strength. Thus to obtain a maximum $\mathcal{I}_1$, we need an applied pulse frequency $\omega_{\mathrm{pulse}}$ such that $\omega_{\mathrm{pulse}}-\epsilon({\bf k})/\hbar\simeq 0$ is satisfied. However, for $\omega_{\mathrm{pulse}}\neq \epsilon(\mathbf{k})/\hbar$\,, $\mathcal{I}_1$ is reduced due to the oscillatory nature of the integrand in Eq.~(\ref{eq:4int1}).
	

	
\begin{figure}[!t]
		\centering
		\includegraphics[width=\linewidth]{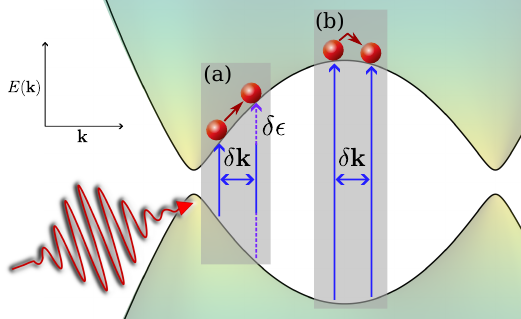}
		\caption{Schematics of energy gap $\epsilon(\mathbf{k})$, showing different regions of energy spectrum. In region (a), corresponding to the steep portion of the spectrum, a small change in $\mathbf{k}$ produces a significant change in $\epsilon(\mathbf{k})$. In contrast, in region (b), where the spectrum is comparatively flat, the same change in $\mathbf{k}$ results in negligible change in $\epsilon(\mathbf{k})$.}
		\label{fig:4scematic}
	\end{figure}

	
	In a time-dependent field, $\mathbf{k}$ changes with time as $\mathbf{k}-\frac{q\vec A(t)}{\hbar}$, hence the above equality {\it may not} be strictly satisfied even for a low intensity pulse with a fixed frequency. In fact, this depends on the curvature of the band spectrum. For a relatively steep region in the band spectrum as shown in Fig.~\ref{fig:4scematic}, a small change in $\mathbf{k}$ in a particular direction evidently results in a relatively large change in $\epsilon(\mathbf{k})$. In contrast, for a relatively flat region in the same band spectrum, the change in momentum does not change $\epsilon ({\bf k})$ substantially. Thus, for a time-varying field, the above equality is best satisfied for a relatively flat region of the band spectrum. The interband coherence term is therefore expected to encode saddle point of the band spectrum in a non-equilibrium set up.

	{\em Response-}
	Next, we shall see that this interesting feature of the interband coherence term is reflected in the response of the system. Since the response is directly linked to the dipole acceleration~\cite{Gaarde_2008} which is proportional to the time derivative of the total current density,  we focus on $\dot{\mathfrak{J}}(t)$ for the rest of this study.
	A straightforward calculation shows that both $x$ and $y$ components of single particle current can be expressed as
	\begin{equation}
	\hspace*{-0.3cm}\mathcal{J}_{x(y)} ({\bf k},t)=\gamma_1({\bf k},t)_{x(y)} n({\bf k},t) +\gamma_2({\bf k},t)_{x(y)}\pi({\bf k},t)\,,
	\end{equation}
	where $\gamma_1,\,\gamma_2$ involves $\epsilon ({\bf k})$, $\nabla_{\bf k}\epsilon ({\bf k})$ and momentum dependent matrix elements of $\mathcal{H}$ (see Supplementary materials).
	For a post-pulse time domain, $n({\bf k},t)$ remains constant (see Eqs.~(\ref{eq:4ng}) and (\ref{eq:4nc})). Thus, the time variation of the {\it post-pulse} current is mainly determined by the $\pi ({\bf k},t)$.
	To confirm these, we next compute interband coherence term and the total current (Eq.~(\ref{eq:total_current})) for an anisotropic Dirac model.
	
\begin{figure}[!t]
		\centering
		\includegraphics[width=\linewidth]{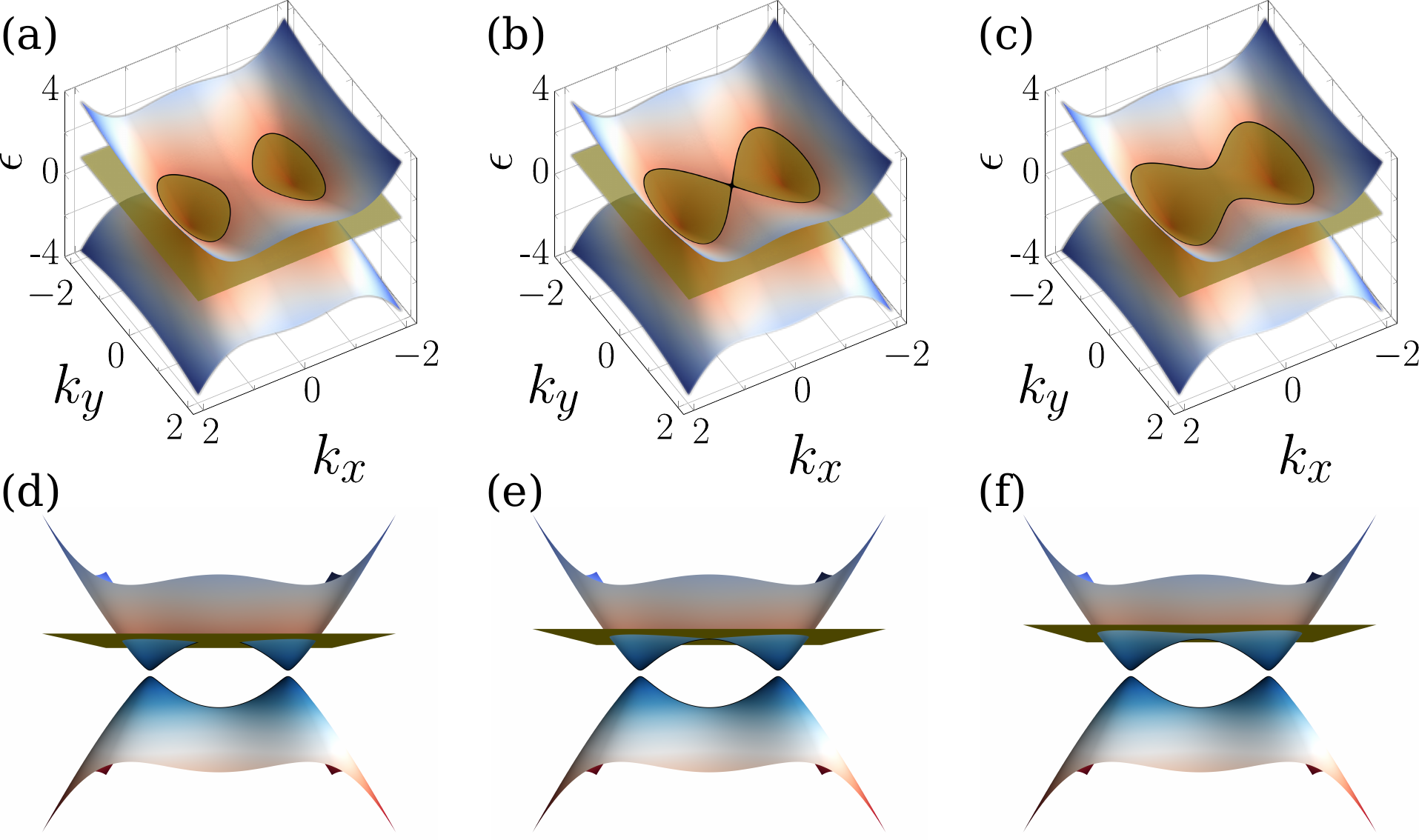}
		\caption{(a)--(c) show 3D view of the FS over the 2D band. As the Fermi level approaches the saddle point, a LT occurs. Panels (d)--(f) display the corresponding cross-sectional view in $k_y=0$ plane.}
		\label{fig:4FS}
	\end{figure}

	{\em Application to Dirac materials--}
	we consider a simple low-energy model Hamiltonian of deformed honeycomb lattice, which can host topological LT. The Hamiltonian reads off
	\begin{equation}
	\mathcal{\hat{H}}(\mathbf{k})=h ({\mathbf{k}})\cdot\sigma,\label{eq:hamiltonian}
	\end{equation}
	where $\sigma_i$'s are the Pauli matrices, $h({\mathbf{k}})=(\alpha\,k_x^2-\delta,\beta\, k_y,m)$, $\alpha$ is the inverse mass, $\beta$ is the velocity and $m$ is the mass of the Dirac fermion. For a fixed $\delta>0$, the variation of $E_F$ can give rise to topological LT. Fig.~\ref{fig:4FS} provides a visual representation of the Fermi surface at different Fermi energies for a fixed $\delta$. Clearly, as the Fermi level traverses the saddle point (Fig.~\ref{fig:4FS}(e)), two distinct regions merge into a single one. For this work, we set $\alpha= 1\, \mathrm{meV}\,\mathrm{\AA}^2, \beta= 1\, \mathrm{meV}\,\mathrm{\AA}$, $\delta=1\, \mathrm{meV}$ and $m=0.1\, \mathrm{meV}$.
	Accordingly, the frequency $\omega$ is set to be in the experimentally feasible $\mathrm{THz}$ range (time in $\mathrm{ps}$) and the electric field is taken to be of the order $\mathrm{MVm^{-1}}$.

\begin{figure}[!t]
		\centering
		\includegraphics[width=\linewidth]{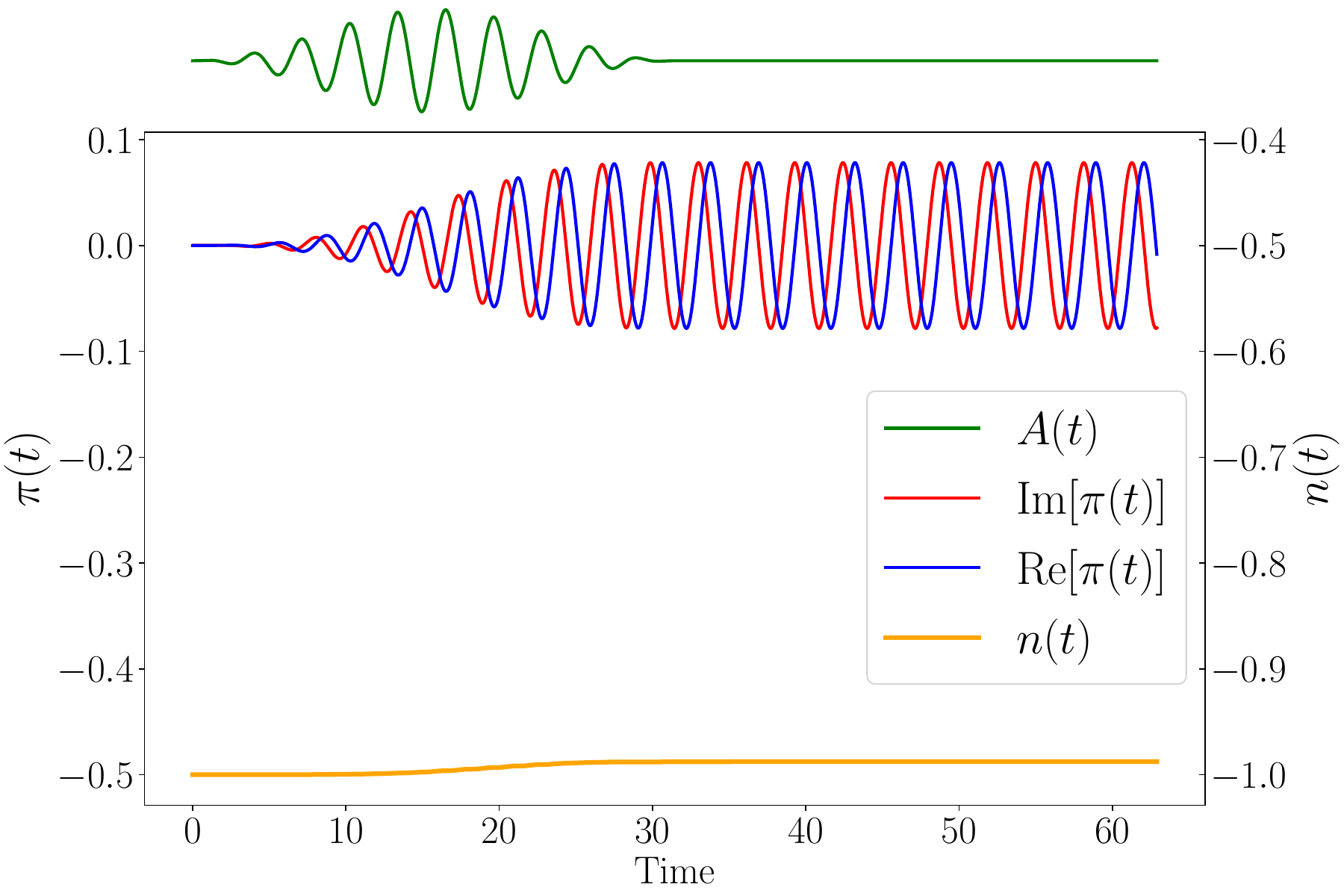}
		\caption{Temporal evolution of interband coherence term ($\pi$) and population difference term ($n$) for the applied field strength $A_0=0.01$ and energy $\hbar\omega_{\mathrm{pulse}}\sim 2$ in appropriate units as discussed in the main text. ${\bf k}$ is set by the equation $\hbar\omega_{\mathrm{pulse}}=\epsilon ({\bf k})$. $\pi$ exhibits post-pulse oscillation, while $n$ remains flat, as clearly seen during the flat portion of the green schematic curve of the applied pulse.}
		\label{fig:4pi-n-0-0.3}
	\end{figure}

	To find $n_g$, $n_e$ and $\pi$ for the Hamiltonian in Eq.~(\ref{eq:hamiltonian}), we numerically solve the LVNE for a $n=10$ -- cycle $\sin^2$ pulse $A_0\sin^2{\left(\omega_{\mathrm{pulse}}t/2n\right)\sin{(\omega_{\mathrm{pulse}}t)}}$ with a particular frequency $\omega_{\mathrm{pulse}}$ with the same initial condition discussed before. Fig.~\ref{fig:4pi-n-0-0.3} illustrates the variation of population difference ($n$) and interband coherence term ($\pi$). The oscillatory interband coherence corroborates the result obtained analytically in Eq.~(\ref{eq:4pi3}). In contrast, $n(t)$ shows negligible time variation in the pulse domain and remains constant in the post-pulse domain, in conjunction with the discussion before.
	
	We next find the response of the system for three distinct cases of Fermi surfaces as shown in Fig.~\ref{fig:4FS}. At $T = 0\,K$, the total current is the cumulative response of all fermions filled up to the Fermi level. For a weak static field, the response is primarily dominated by the charged particles near the Fermi sea due to available states with momentum $\mathbf{k}_F-\frac{q \vec{A}}{\hbar}$, where $\vec{A}$ is the static gauge potential. In contrast, for the time-varying weak field with a frequency $\omega_{\rm pulse}$, the valence electrons can resonantly couples with the available electronic states near the Fermi surface, satisfying the relation $\hbar\omega\equiv \epsilon(\mathbf{k}_F-\frac{q \vec{A}(t)}{\hbar})$. Thus the current is primarily dominated by interband contribution. On the contrary, the intraband current contribution from the very high energy ($E_e({\bf k})>E_F$) states is negligible for a weak field.

	\begin{figure}[!b]
		\centering
		\includegraphics[width=\linewidth]{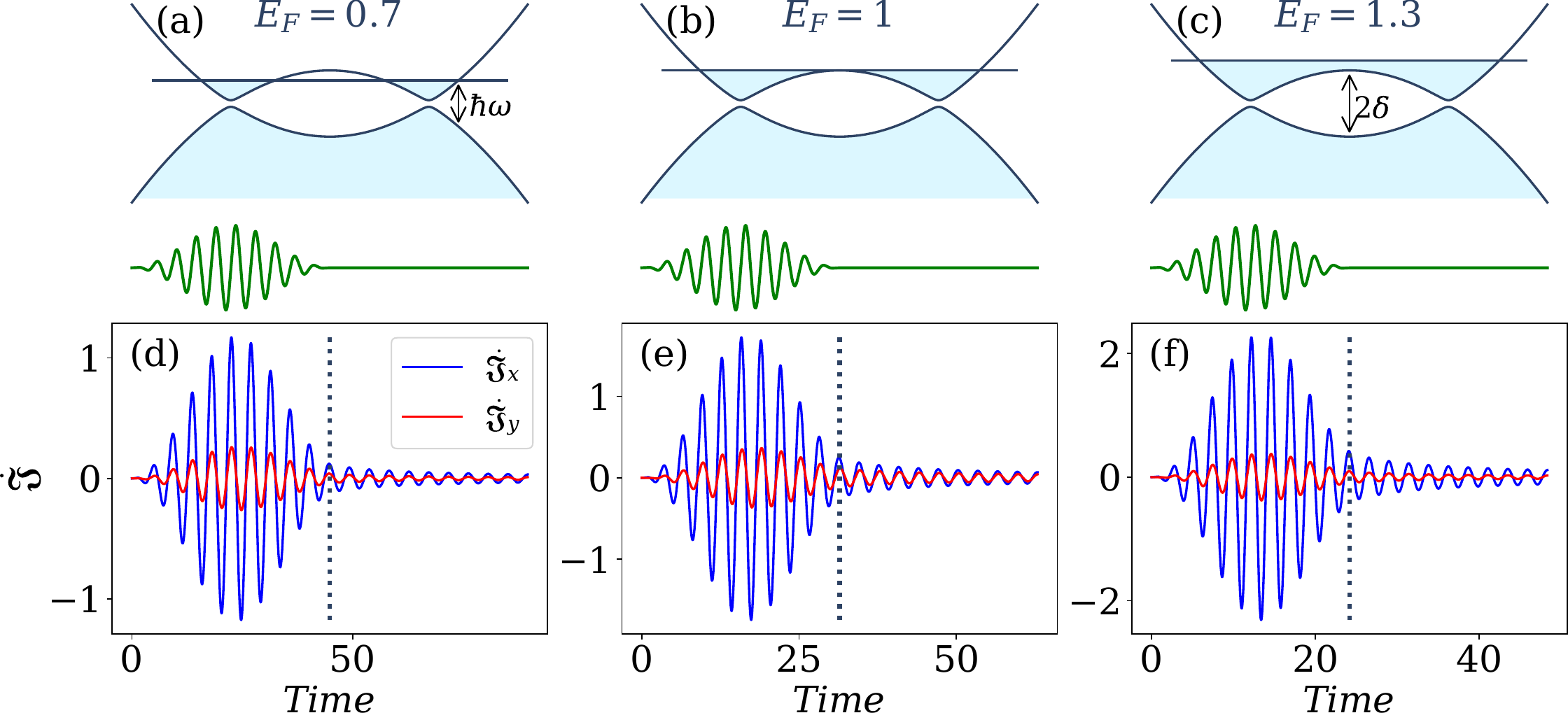}
		\caption{Response of the system for different Fermi energies. (a)-(c) represent the Fermi level positions, starting below the saddle point and moving upward, with LT occurring at (b). (d)-(f) depict the corresponding responses. The post-pulse oscillations at the end of the applied pulse (green) vary depending on the position of Fermi level. Vertical dotted lines indicate the end of the applied pulse.}
		\label{fig:particle_current}
	\end{figure}
	
	\begin{figure}[!t]
		\centering
		\includegraphics[width=\linewidth]{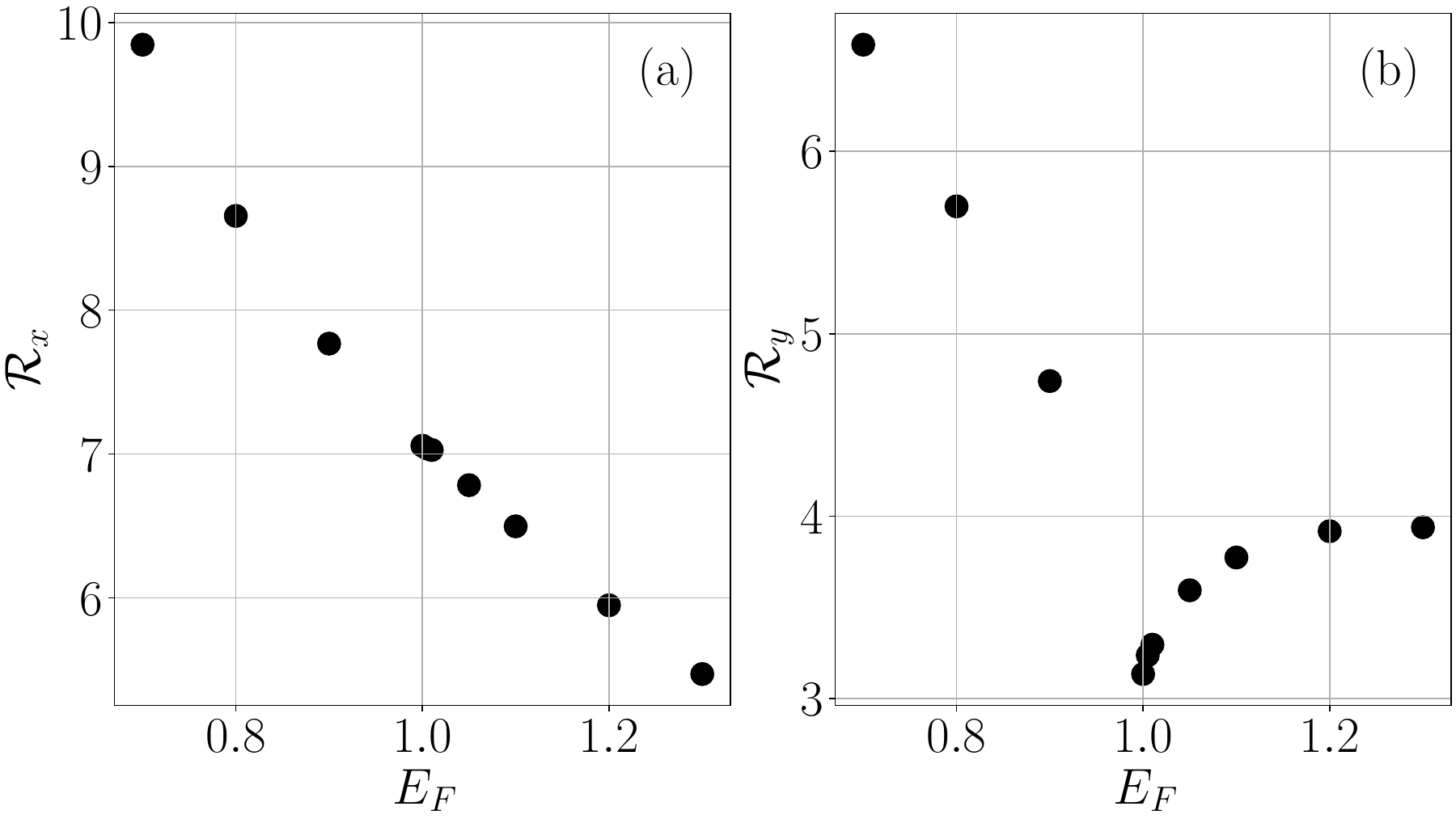}
		\caption{Variation of $\mathcal{R}_x$ and $\mathcal{R}_y$ with Fermi level $E_F$. (a) $\mathcal{R}_x$ monotonically decreases with $E_F$ as the band-curvature along $k_x$ decreases gradually around the FS. (b) In contrast, $\mathcal{R}_y$ shows a minimum  since the band curvature along $k_y$ exhibits an abrupt change at the saddle point.}
		\label{fig:4RS}
	\end{figure}
	
	Figure~\ref{fig:particle_current}(a) refers to the case when $E_F$ is below the saddle point and the incident light frequency resonantly couples valence electrons with the electrons on the Fermi surface. In this case, we find post-pulse oscillation as a manifestation of interband coherent transition as shown in Fig.~\ref{fig:particle_current}(d). The oscillation amplitude differs along $x$ from $y$ due to the anisotropic nature of the Fermi contour as will be discussed shortly. Fig.~\ref{fig:particle_current}(b) and (c) refer to the cases when $E_F$ just touch the saddle point and far above the saddle point, respectively. In both cases, we find post-pulse oscillation as illustrated in Fig.~\ref{fig:particle_current}(e) and (f), respectively, when light pulse resonantly couples both the valence and conduction electrons.
	
	As discussed previously, the amplitude of the post-pulse oscillation, i.e. $\mathcal{I}_1$ in Eq.~(\ref{eq:4int1}) serves as a key indicator of the saddle point in the energy spectrum. However, the overall response and the magnitude of oscillation also depends on $E_F$. As we increase $E_F$, the total current enhances as the number of carriers increases. This is evident in Fig.~\ref{fig:particle_current}. 
	Thus, to assess whether the post-pulse oscillation amplitude increases near the saddle point of the Fermi level, we compare it to the maximum amplitude during the pulse for different Fermi level. To quantify this, we define a quantity $\mathcal{R}_i$, taking the ratio between the maximum magnitude of the response during pulse duration with the magnitude of post-pulse oscillation, 
	\begin{equation}
	\mathcal{R}_i=\frac{\mathrm{max}(|\mathfrak{\dot{J}}_i(\mathrm{during~pulse})|)}{|\mathfrak{\dot{J}}_i(\mathrm{end~of~pulse})|} 
	\end{equation}
	with $i\in(x,y)$. Fig.~\ref{fig:4RS}(a,b) illustrates this behavior both in the $x$ and $y$ directions as we vary $E_F$. While $\mathcal{R}_x$ decreases with $E_F$, $\mathcal{R}_y$ shows a minimum value as $E_F$ touches the saddle point of the band spectrum. Thus the minimum value of $R_y$ serves as a probe of topological LT.
	
\begin{figure}
		\centering
		\includegraphics[width=\linewidth]{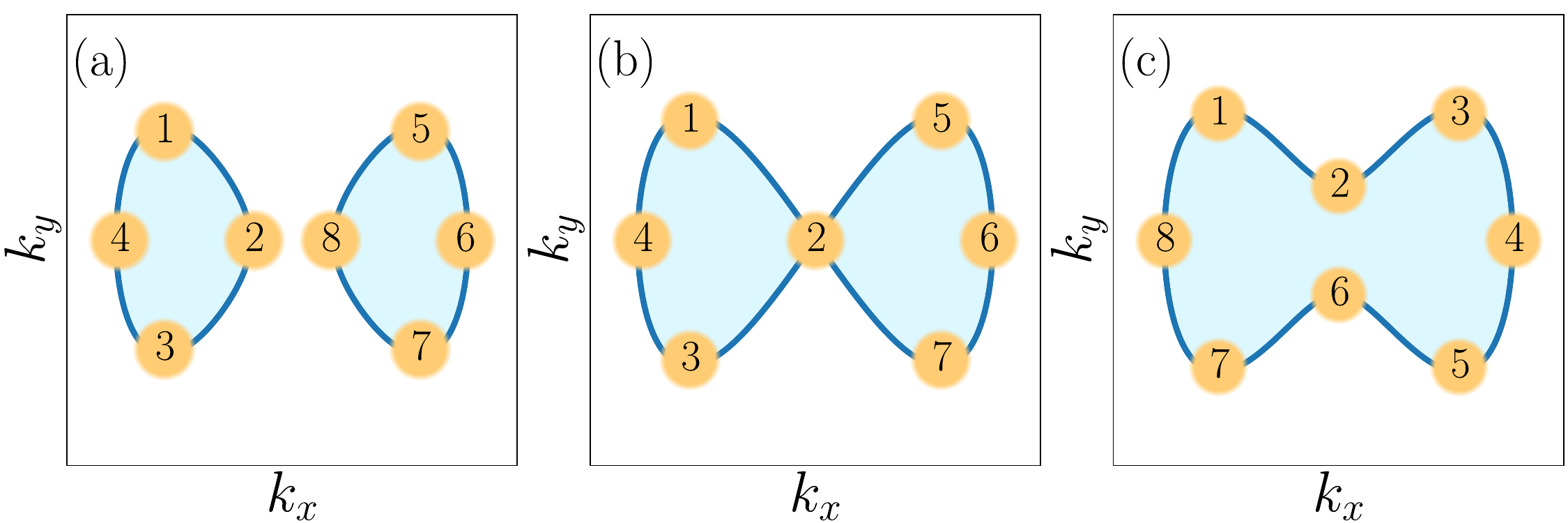}
		\caption{Plot of Fermi surfaces in $k_x-k_y$ plane for three distinct cases as shown in Fig.~(\ref{fig:4FS}). (a) For the Fermi level below the saddle point, $\frac{\partial\epsilon(\mathbf{k})}{\partial k_x}\sim 0$ at $1$, $3$, $5$, $7$ points and $\frac{\partial\epsilon(\mathbf{k})}{\partial k_y}\sim0$ at $2$, $4$, $6$, $8$ points. As the Fermi level approaches the saddle point, the region over which $\epsilon(\mathbf{k})$ is invariant in a given direction expands and Fermi level at the saddle point, two Fermi pockets merge into one (b), with point number $2$ exhibiting invariance along both directions. Above the saddle points (c), points $2$ and $6$ remain invariant only along $k_x$ not $k_y$.}
		\label{fig:4LT}
	\end{figure}

	The difference in $\mathcal{R}_x$ and $\mathcal{R}_y$ can be understood by the anisotropic shape of the Fermi contour as shown in Fig.~\ref{fig:4LT}. The eight marked points in Fig.~\ref{fig:4LT}(a) indicate zero gradient in specific directions, where change in $k_x$ or $k_y$ do not change $\epsilon(\mathbf{k})$, thus maximizing $\mathcal{I}_1$ in Eq.~(\ref{eq:4int1}). At $1$, $3$, $5$, and $7$ points $\frac{\partial \epsilon(\mathbf{k})}{\partial k_x} \sim 0$, while at $2$, $4$, $6$, and $8$ points $\frac{\partial \epsilon(\mathbf{k})}{\partial k_y} \sim 0$. This indicate that the Fermi contour in invariant along $k_x$ at four points and along $k_y$ for other four points. As the Fermi level rises towards the saddle point, the expanding Fermi surface enlarges these invariant regions around the aforementioned points, enhancing the post-pulse response amplitude and thereby reducing $\mathcal{R}_x$ and $\mathcal{R}_y$. Above the LT, at points $2$ and $6$ in Fig.~\ref{fig:4LT}(c), $\epsilon(\mathbf{k})$ loses its invariance along $k_y$ but remains invariant along $k_x$. As the Fermi level rises--starting from below the saddle point, crossing the LT, and continuing upward--the invariant regions of the Fermi surfaces initially expand. Beyond the transition, the invariant regions along $k_x$ continue to grow with the expanding Fermi surface, while those along $k_y$ diminish abruptly. As a results, this behavior is reflected in a gradual decrease  in $\mathcal{R}_x$ (Fig.~\ref{fig:4RS}(a)), and a  sharp change in $\mathcal{R}_y$ (Fig.~\ref{fig:4RS}(b)), clearly marking the Lifshitz transition point.

\begin{figure}
		\centering
		\includegraphics[width=\linewidth]{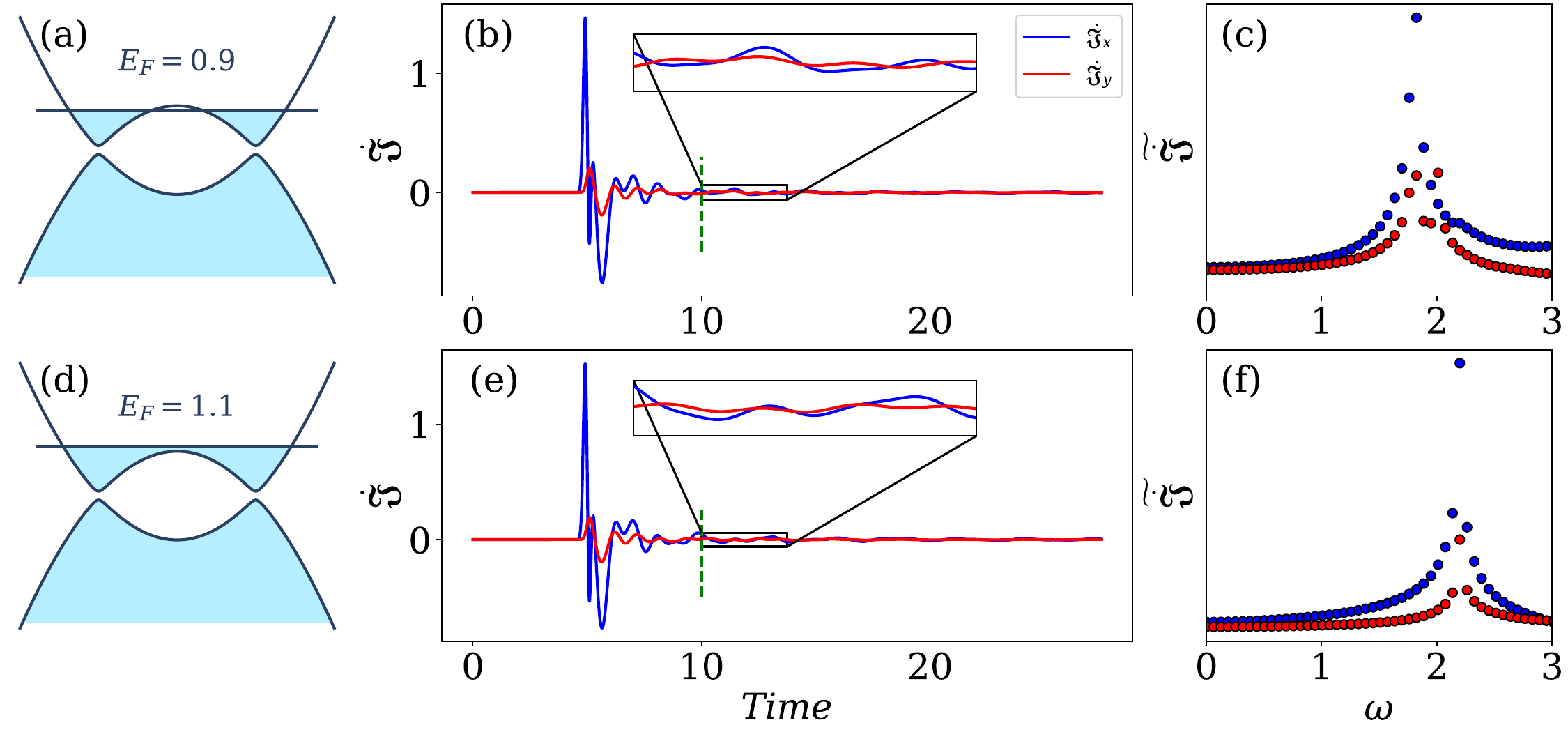}
		\caption{(a) and (d) show cross-sectional views of the band at the $k_y=0$ plane at different Fermi levels. (b) and (e) shows the pulse dynamics highlighting the post-pulse oscillations (zoomed insets) for the applied Gaussian pulse. (c) and (f) display the Fourier components of the post-pulse oscillation (in arbitrary units), with peaks at $\hbar\omega \simeq 2E_F$ identifying the Fermi level.}
		\label{fig:4gaussianPulse}
	\end{figure}

	{\em Application of Gaussian pulse: Location of Fermi surface.-}
	To this end, we discuss how to determine appropriate frequency of the $\sin^2$ pulse in a generic two band model with a finite gap so that the method discussed above can be applied efficiently. In doing so, we initially apply a Gaussian pulse $\vec{A}(t)=\vec{A}_0e^{-\frac{t^2}{2\sigma^2}}$ of low amplitude, to the system. The standard deviation
	parameter $\sigma$ of the pulse is chosen such that in frequency space it should effectively contain all the frequency component within which we expect the Fermi level to be present in the system. Fig.~\ref{fig:4gaussianPulse} demonstrates that the low amplitude Gaussian pulse identifies the energy of the Fermi level for proper choice of $\sigma$. In frequency domain, the post-pulse oscillation shows a peak in the frequency where $\hbar\omega\simeq 2E_F$ (Fig.~\ref{fig:4gaussianPulse}(c) and Fig.~\ref{fig:4gaussianPulse}(f)). This frequency $\omega$ can now be used as the  frequency, $\omega_{\mathrm{pulse}}$ of the $\sin^2$ pulse.

	{\em Conclusion.-}
	In conclusion,  we introduce a method to identify Lishitz transitions in the Fermi surface through non-equilibrium light-matter interactions. We find that when the Fermi surface is near the saddle point of the band spectrum, the relative amplitude between the peak current during the pulse and that of the current during the post-pulse reaches a minimum, unlike cases where the Fermi energy lies below or above the saddle point.  This behavior arises from the interband coherence term in the current, which couples the valence and conduction electrons resonantly.
	We further demonstrate that a Gaussian pulse can be effectively employed to locate the Fermi surface, hence enabling us to precisely select the frequency of the $\sin^2$ pulse.
	In sum, our results enable real-time observation of change in the FS topology during dynamic processes and provides a simple, non-destructive technique. The non-destructive nature allows the same sample to be reused, tuning the FS near the singular point repeatedly. This facilitates the study of bulk properties in addition to surface properties, making the approach highly versatile and practical for experimental applications.
	\section{acknowledgment}
	We acknowledge financial support from the Department of Atomic Energy (DAE), Govt. of India, through the project Basic Research in Physical and Multidisciplinary Sciences via RIN4001. DD acknowledges the Virgo cluster, where most of the numerical calculations were performed. KS acknowledges funding from the Science and Engineering Research Board (SERB) under SERB-MATRICS Grant No. MTR/2023/000743.

\onecolumngrid
\section*{Supplementary Material: Frequency-selective amplification of nonlinear response in strongly correlated bosons}

This Supplementary Material provides further details on the Semiconductor Bloch Equation and the single particle current expression. In the first part, we introduce the dipole matrix element and then derive the Semiconductor Bloch Equations in velocity gauge. In the second part, we derive the single-particle current expression Eq.~(10) of the main text.

\section{Semiconductor Bloch Equation in Different Gauges}\label{sec:SBE}

In periodic crystals, the eigenstates are Bloch eigenstates $\Phi_{m,\mathbf{k}}(\mathbf{x})=u_{m,\mathbf{k}}(\mathbf{x})e^{i\mathbf{k}.\mathbf{x}}$, with eigenenergy $E_m(\mathbf{k})$. To determine the response of the system, we first compute the elements of the dipole transition matrix, denoted as $\langle\Phi_{m',k'}|\hat{\mathbf{x}}|\Phi_{m,k}\rangle$. To obtain this dipole matrix element we proceed as follows
\begin{align}
    &\frac{\partial}{\partial \mathbf{k}}\int \Phi_{m',\mathbf{k'}}^*(\mathbf{x})\Phi_{m,\mathbf{k}}(\mathbf{x})\,d\mathbf{x}\nonumber\\
    =&\frac{\partial}{\partial \mathbf{k}}\int u_{m',\mathbf{k'}}^*(\mathbf{x})u_{m,\mathbf{k}}(\mathbf{x})e^{i(\mathbf{k}-\mathbf{k'}).\mathbf{x}}\,d\mathbf{x}\nonumber\\
    =&\int u_{m',\mathbf{k'}}^*(\mathbf{x})\left\{\partial_{\mathbf{k}}u_{m,\mathbf{k}}(\mathbf{x})e^{i(\mathbf{k}-\mathbf{k'}).\mathbf{x}}+u_{m,\mathbf{k}}(\mathbf{x})e^{i(\mathbf{k}-\mathbf{k'}).\mathbf{x}}i\mathbf{x}\right\}\,d\mathbf{x}\nonumber\\
    =&\int u_{m',\mathbf{\mathbf{k'}}}^*(\mathbf{x})\partial_{\mathbf{k}}u_{m,\mathbf{k}}(\mathbf{x})e^{i(\mathbf{k}-\mathbf{k'})\mathbf{x}}\,dx+\int i\mathbf{x}u_{m',\mathbf{k'}}^*(\mathbf{x})u_{m,\mathbf{k}}(\mathbf{x})e^{i(\mathbf{k}-\mathbf{k'})\mathbf{x}}\,d\mathbf{x}\,.\nonumber
\end{align}
This can further be rewritten as
\begin{align}
    &i\int u_{m',\mathbf{k'}}^*(\mathbf{x})\mathbf{x}u_{m,\mathbf{k}}(\mathbf{x})e^{i(\mathbf{k}-\mathbf{k'})\mathbf{x}}\,dx=\frac{\partial}{\partial \mathbf{k}}\int \Phi_{m',\mathbf{k'}}^*(\mathbf{x})\Phi_{m,\mathbf{k}}(\mathbf{x})\,d\mathbf{x}\nonumber\\
    &~~~~~~~~~~~~~~~~~~~~~~~~~~~~~~~~~~~~~~~~~~~~-\int u_{m',\mathbf{k'}}^*(\mathbf{x})\partial_ku_{m,\mathbf{k}}(\mathbf{x})e^{i(\mathbf{k}-\mathbf{k'})\mathbf{x}}\,d\mathbf{x}\nonumber\\
    \Rightarrow ~&i\int u_{m,\mathbf{k}}^*(\mathbf{x})\mathbf{x}u_{m',\mathbf{k'}}(\mathbf{x})e^{i(\mathbf{k'}-\mathbf{k})\mathbf{x}}\,dx=\frac{\partial}{\partial \mathbf{k}}\sum_{R}e^{(\mathbf{k}-\mathbf{k'})R}\int_{BZ} u_{m',\mathbf{k'}}^*(\mathbf{x})u_{m,\mathbf{k}}(\mathbf{x})e^{i(\mathbf{k}-\mathbf{k'})\mathbf{x}}\,d\mathbf{x}\nonumber\\
    &~~~~~~~~~~~~~~~~~~~~~~~~~~~~~~~~~~~~~~~~~~~~-\sum_{R}e^{i(\mathbf{k}-\mathbf{k'})R}\int_{BZ} u_{m',\mathbf{k'}}^*(\mathbf{x})\partial_ku_{m,\mathbf{k}}(\mathbf{x})e^{i(\mathbf{k}-\mathbf{k'})\mathbf{x}}\,d\mathbf{x}\nonumber\\
    \Rightarrow ~&\langle\Phi_{m',\mathbf{k'}}|\hat{\mathbf{x}}|\Phi_{m,\mathbf{k}}\rangle=-i\delta_{m,m'}\frac{\partial}{\partial \mathbf{k}}\delta(\mathbf{k}-\mathbf{k'})+\vec{d}_{m',m}(\mathbf{k})\delta(\mathbf{k}-\mathbf{k'})\label{eq:dipole_element}\,,
\end{align}
where
\begin{equation}
  \vec{d}_{m',m}(\mathbf{k})=i\int u_{m',\mathbf{k}}^*(\mathbf{\mathbf{x}})\partial_{\mathbf{k}} u_{m,\mathbf{k}}(\mathbf{x})\,d\mathbf{x}\,.
\end{equation}
The diagonal elements of $\vec d$ give the Berry connection of the band, $\vec{\zeta}_m(\mathbf{k})=\vec{d}_{m,m}(\mathbf{k})$, and the curl of this quantity gives rise to the gauge invariant quantity Berry curvature $\vec{\Omega}_m(\mathbf{k})=\nabla\times \vec{\zeta}_m(\mathbf{k})$.

For a generic statistical mixture of states $|\Psi\rangle=\sum_jp_j|\Psi_j\rangle$, the density matrix is defined as:
\begin{equation}
    \hat{\rho}=\sum_jp_j|\Psi_j\rangle\langle\Psi_j|\,.
\end{equation}
For pure states $|\Psi_j\rangle$, i.e. $p_j=1$, we can get
\begin{align}
    \hat{\rho}&=\sum_{\mathbf{k},\mathbf{k'}}\sum_{m,m'}a_m(\mathbf{k},t)a_{m'}^*(\mathbf{k'},t)|\Phi_{m,k}\rangle\langle\Phi_{m',k'}|\\
    &=\sum_{\mathbf{k},\mathbf{k'}}\sum_{m,m'}\rho_{m,m'}(\mathbf{k},\mathbf{k'},t)|\Phi_{m,\mathbf{k}}\rangle\langle\Phi_{m',\mathbf{k'}}|\,.\label{eq:density_matrix}
\end{align}
Together with Eq.~\ref{eq:density_matrix} and LVNE in Eq.~3 of main text, we find
\begin{align}
    i\hbar\sum_{\mathbf{k},\mathbf{k'}}\sum_{m,m'}\dot{\rho}_{m,m'}(\mathbf{k},\mathbf{k'},t)|\Phi_{m,\mathbf{k}}&\rangle\langle\Phi_{m',\mathbf{k'}}|=\sum_{\mathbf{k},\mathbf{k'}}\sum_{m,m'}\rho_{m,m'}(\mathbf{k},\mathbf{k'},t)\hat{\mathcal{H}}(\mathbf{k})|\Phi_{m,\mathbf{k}}\rangle\langle\Phi_{m',\mathbf{k'}}|\nonumber\\
    &-\sum_{\mathbf{k},\mathbf{k'}}\sum_{m,m'}\rho_{m,m'}(\mathbf{k},\mathbf{k'},t)|\Phi_{m,\mathbf{k}}\rangle\langle\Phi_{m',\mathbf{k'}}|\hat{\mathcal{H}}(\mathbf{k})\label{eq:dif_dm}\,.
\end{align}
In velocity gauge $\mathbf{k}\to\mathbf{k}-\frac{q}{\hbar}\vec{A}(t)$. In this case, the time derivative will be
\begin{equation}
    \frac{d}{dt}\to\dot{\mathbf{k}}(t).\frac{\partial}{\partial \mathbf{k}}+\frac{\partial}{\partial t}\equiv \frac{q}{\hbar}\vec{E}(t).\frac{\partial}{\partial \mathbf{k}}+\frac{\partial}{\partial t}\,.
\end{equation}
From Eq.~\ref{eq:dif_dm}, we now obtain the diagonal terms of the density matrix as
\begin{align}
    &i\hbar\dot{\rho}_{g,g}(\mathbf{k},t)=\rho_{e,g}(\mathbf{k},t)\langle\Phi_g(\mathbf{k},t)|\mathcal{\hat{H}}(\mathbf{k},t)|\Phi_e(\mathbf{k},t)\rangle-\rho_{g,e}(\mathbf{k},t)\langle\Phi_e(\mathbf{k},t)|\mathcal{\hat{H}}(\mathbf{k},t)|\Phi_g(\mathbf{k},t)\rangle\nonumber\\
    \Rightarrow~&\dot{\rho}_{g,g}(\mathbf{k},t)=\rho_{e,g}(\mathbf{k},t)\langle\Phi_g(\mathbf{k},t)|\dot{\Phi}_e(\mathbf{k},t)\rangle-\rho_{g,e}(\mathbf{k},t)\langle\Phi_e(\mathbf{k},t)|\dot{\Phi}_g(\mathbf{k},t)\rangle\nonumber\\
    \Rightarrow~&\dot{\rho}_{g,g}(\mathbf{k},t)=\rho_{e,g}(\mathbf{k},t)\langle\Phi_g(\mathbf{k},t)|\frac{q}{\hbar}\vec{E}(t)\frac{\partial}{\partial\mathbf{k}}\Phi_e(\mathbf{k},t)\rangle+\rho_{e,g}(\mathbf{k},t)\langle\Phi_g(\mathbf{k},t)|\frac{\partial}{\partial t}\Phi_e(\mathbf{k},t)\rangle\nonumber\\
    &-\rho_{g,e}(\mathbf{k},t)\langle\Phi_e(\mathbf{k},t)|\frac{q}{\hbar}\vec{E}(t)\frac{\partial}{\partial\mathbf{k}}\Phi_g(\mathbf{k},t)\rangle-\rho_{g,e}(\mathbf{k},t)\langle\Phi_e(\mathbf{k},t)|\frac{\partial}{\partial t}\Phi_g(\mathbf{k},t)\rangle\nonumber\nonumber\\
    \Rightarrow~&\hbar\dot{\rho}_{g,g}(\mathbf{k},t)=-iq\vec{E}(t)\left[\rho_{e,g}(\mathbf{k},t)\vec{d}_{g,e}(\mathbf{k},t)-\rho_{g,e}(\mathbf{k},t)\vec{d}_{e,g}(\mathbf{k},t)\right]\,.\label{eq:ng-v}
\end{align}
Similarly,
\begin{equation}
    \hbar\dot{\rho}_{e,e}(\mathbf{k},t)=-iq\vec{E}(t)\left[\rho_{g,e}(\mathbf{k},t)\vec{d}_{e,g}(\mathbf{k},t)-\rho_{e,g}(\mathbf{k},t)\vec{d}_{g,e}(\mathbf{k},t)\right]\,.\label{eq:ne-v}
\end{equation}
The off-diagonal terms are 
\begin{align}
    &\hbar\dot{\rho}_{e,g}(\mathbf{k},t)=\rho_{g,g}(\mathbf{k},t)\langle\Phi_e(\mathbf{k},t)|\mathcal{\hat{H}}(\mathbf{k},t)|\Phi_g(\mathbf{k},t)\rangle+\rho_{e,g}(\mathbf{k},t)\langle\Phi_e(\mathbf{k},t)|\mathcal{\hat{H}}(\mathbf{k},t)|\Phi_e(\mathbf{k},t)\rangle\nonumber\\
    &-\rho_{e,g}(\mathbf{k},t)\langle\Phi_g(\mathbf{k},t)|\mathcal{\hat{H}}(\mathbf{k},t)|\Phi_g(\mathbf{k},t)\rangle-\rho_{e,e}(\mathbf{k},t)\langle\Phi_e(\mathbf{k},t)|\mathcal{\hat{H}}(\mathbf{k},t)|\Phi_g(\mathbf{k},t)\rangle\nonumber\\
    \Rightarrow~&\dot{\rho}_{e,g}(\mathbf{k},t)=\rho_{g,g}(\mathbf{k},t)\langle\Phi_e(\mathbf{k},t)|\dot{\Phi}_g(\mathbf{k},t)\rangle+\rho_{e,g}(\mathbf{k},t)\langle\Phi_e(\mathbf{k},t)|\dot{\Phi}_e(\mathbf{k},t)\rangle\nonumber\\
    &-\rho_{e,g}(\mathbf{k},t)\langle\Phi_g(\mathbf{k},t)|\dot{\Phi}_g(\mathbf{k},t)\rangle-\rho_{e,e}(\mathbf{k},t)\langle\Phi_e(\mathbf{k},t)|\dot{\Phi}_g(\mathbf{k},t)\rangle\nonumber\\
    \Rightarrow~&\dot{\rho}_{e,g}(\mathbf{k},t)=\rho_{g,g}(\mathbf{k},t)\langle\Phi_e(\mathbf{k},t)|\frac{q}{\hbar}\vec{E}(t)\frac{\partial}{\partial\mathbf{k}}\Phi_g(\mathbf{k},t)\rangle+\rho_{e,g}(\mathbf{k},t)\langle\Phi_e(\mathbf{k},t)|\frac{q}{\hbar}\vec{E}(t)\frac{\partial}{\partial\mathbf{k}}\Phi_e(\mathbf{k},t)\rangle\nonumber\\
    &+\rho_{e,g}(\mathbf{k},t)\langle\Phi_e(\mathbf{k},t)|\frac{\partial}{\partial t}\Phi_e(\mathbf{k},t)\rangle\nonumber-\rho_{e,g}(\mathbf{k},t)\langle\Phi_g(\mathbf{k},t)|\frac{q}{\hbar}\vec{E}(t)\frac{\partial}{\partial \mathbf{k}}\Phi_g(\mathbf{k},t)\rangle\\
    &-\rho_{e,g}(\mathbf{k},t)\langle\Phi_g(\mathbf{k},t)|\frac{\partial}{\partial t}\Phi_g(\mathbf{k},t)\rangle\nonumber-\rho_{e,e}(\mathbf{k},t)\langle\Phi_e(\mathbf{k},t)|\frac{q}{\hbar}\vec{E}(t)\frac{\partial}{\partial \mathbf{k}}\Phi_g(\mathbf{k},t)\rangle\nonumber\\
    \Rightarrow~&\hbar\dot{\rho}_{e,g}(\mathbf{k},t)=-i\left\{\left(E_e(\mathbf{k},t)-E_g(\mathbf{k},t)\right)+q\vec{E}(t).\left(\vec{d}_{e,e}(\mathbf{k},t)-\vec{d}_{g,g}(\mathbf{k},t)\right)\right\}\rho_{e,g}(\mathbf{k},t)\nonumber\\
    &+iq\vec{E}(t).\vec{d}_{e,g}\left(\rho_{e,e}(\mathbf{k},t)-\rho_{g,g}(\mathbf{k},t)\right)\,.\label{eq:pi-v}
\end{align}
As explained in the main text, we interpret the diagonal elements of the density matrix as population terms ($n_m=\rho_{m,m}$) and the off-diagonal element as the interband coherence term ($\pi=\rho_{e,g}$). With these definitions, Eqs.~(\ref{eq:ng-v}), (\ref{eq:ne-v}), and (\ref{eq:pi-v}) yield Eqs. (4), (5), and (6) in the main text.


\section{Single particle current}
To understand the dynamics of charged carriers, we first compute the single particle current operator as
\begin{equation}
    \hat{\mathcal{J}}(\mathbf{k},t)=\frac{1}{\hbar}\nabla_{\mathbf{k}}\hat{\mathcal{H}}(\mathbf{k},t)\,.
\end{equation}
The expectation value of the current can be simply obtained as
\begin{equation}
    \mathcal{J}(\mathbf{k},t)=\frac{q}{\hbar}\mathrm{Tr}\left(\hat{\rho}(\mathbf{k},t)\nabla_{\mathbf{k}}\hat{\mathcal{H}}(\mathbf{k},t)\right)\,.\label{eq:expectation-current}
\end{equation}
For our case, the Hamiltonian has the form
\begin{equation}
\mathcal{\hat{H}}(\mathbf{k})=h ({\mathbf{k}})\cdot\sigma\,,\label{eq:hamiltonian}
\end{equation}
and any generalized state can be written as $|\Psi(\mathbf{k},t)\rangle=\mathcal{A}(\mathbf{k},t)|\psi_g(\mathbf{k},t)\rangle+\mathcal{B}(\mathbf{k},t)|\psi_e(\mathbf{k},t)\rangle$,
where $|\psi_g(\mathbf{k},t)\rangle$ and $|\psi_e(\mathbf{k},t)\rangle$ refer to the state associated with the valence and conduction band respectively.
We next define population of valence and conduction bands as 
\begin{align}
n_g(\mathbf{k},t)&=|\mathcal{A}(\mathbf{k},t)|^2\,,\\
n_e(\mathbf{k},t)&=|\mathcal{B}(\mathbf{k},t)|^2
\end{align}
respectively and the interband coherence term as
\begin{equation}
\pi(\mathbf{k},t)=\mathcal{A}(\mathbf{k},t)\mathcal{B}^*(\mathbf{k},t)\,.
\end{equation}
The first order differential equations of these terms can be obtained by solving the time-dependent Schr\"odinger equation
\begin{equation}
i\hbar\frac{d |\Psi(\mathbf{k},t)\rangle}{d t}=\mathcal{\hat{H}}|\Psi(\mathbf{k},t)\rangle\label{eq:sup-ham}\,.
\end{equation}
To calculate $\mathcal{J}(\mathbf{k},t)$ in Eq.~\ref{eq:expectation-current}, we now define density matrix $\hat{\rho}(\mathbf{k},t)$ as
\begin{align}
\hat{\rho}(\mathbf{k},t)=|\Psi(\mathbf{k},t)\rangle\langle\Psi(\mathbf{k},t)|\,.
\end{align}
The matrix representation of $\hat \rho$ in $|\psi_g\rangle$ and $|\psi_e\rangle$ basis is
\begin{align}
\hat{\rho}(\mathbf{k},t)=\begin{pmatrix}
|\mathcal{A}(\mathbf{k},t)|^2 && \mathcal{A}^*(\mathbf{k},t)\mathcal{B}(\mathbf{k},t) \\
\mathcal{B}^*(\mathbf{k},t)\mathcal{A}(\mathbf{k},t) && |\mathcal{B}(\mathbf{k},t)|^2
\end{pmatrix}.
\end{align}
The Hamiltonian becomes diagonal in its eigenbasis $|\psi_g\rangle$ and $|\psi_e\rangle$, which is achieved via a similarity transformation $\hat{\mathcal{U}}^{\dagger}\hat{\mathcal{H}}\hat{\mathcal{U}}\equiv \hat{\Lambda}_d$, where $d$ denotes diagonal.


We next express current operator $\hat{\mathcal{J}}(\mathbf{k},t)$ in the eigenbasis of the Hamiltonian. This leads the transformation of the Pauli matrices as
\begin{align}
\mathcal{U}^{\dagger}\sigma_x\mathcal{U}&=\left(
\begin{array}{cc}
-\frac{h_x}{\sqrt{h_x^2+h_y^2+h_z^2}} & \frac{h_x h_z-i h_y \sqrt{h_x^2+h_y^2+h_z^2}}{\sqrt{\left(h_x^2+h_y^2\right) \left(h_x^2+h_y^2+h_z^2\right)}} \\
\frac{h_x h_z+i h_y \sqrt{h_x^2+h_y^2+h_z^2}}{\sqrt{\left(h_x^2+h_y^2\right) \left(h_x^2+h_y^2+h_z^2\right)}} & \frac{h_x}{\sqrt{h_x^2+h_y^2+h_z^2}} \\
\end{array}
\right)\,,\\
\mathcal{U}^{\dagger}\sigma_y\mathcal{U}&=\left(
\begin{array}{cc}
-\frac{h_y}{\sqrt{h_x^2+h_y^2+h_z^2}} & \frac{h_y h_z+i h_x \sqrt{h_x^2+h_y^2+h_z^2}}{\sqrt{\left(h_x^2+h_y^2\right) \left(h_x^2+h_y^2+h_z^2\right)}} \\
\frac{h_y h_z-i h_x \sqrt{h_x^2+h_y^2+h_z^2}}{\sqrt{\left(h_x^2+h_y^2\right) \left(h_x^2+h_y^2+h_z^2\right)}} & \frac{h_y}{\sqrt{h_x^2+h_y^2+h_z^2}} \\
\end{array}
\right)\,.
\end{align}
With this, the current in Eq.~\ref{eq:expectation-current} is found to be
\begin{align}
&\mathcal{J}_x(\mathbf{k},t)=q\alpha\left(k_x-\frac{qA(t)}{\hbar}\right)\left[\frac{h_x}{\sqrt{h_x^2+h_y^2+h_z^2}}n(\mathbf{k},t)+2\,{\mathrm{Re}}\left(\pi(\mathbf{k},t)\frac{h_x h_z-i h_y \sqrt{h_x^2+h_y^2+h_z^2}}{\sqrt{\left(h_x^2+h_y^2\right) \left(h_x^2+h_y^2+h_z^2\right)}}\right)\right]\,,\nonumber\\
&\mathcal{J}_y(\mathbf{k},t)=q\beta\left[\frac{h_y}{\sqrt{h_x^2+h_y^2+h_z^2}}n(\mathbf{k},t)+2\,{\mathrm{Re}}\left(\pi(\mathbf{k},t)\frac{h_y h_z+i h_x \sqrt{h_x^2+h_y^2+h_z^2}}{\sqrt{\left(h_x^2+h_y^2\right) \left(h_x^2+h_y^2+h_z^2\right)}}\right)\right]\label{eq:jy}\nonumber\,,
\end{align}
where $h_x=\alpha k_x^2-\delta$, $h_y=\beta k_y$ and $h_z=m$. Evidently, the single particle current contains the population difference term $n(\mathbf{k},t)$ and interband coherence term $\pi(\mathbf{k},t)$.

\vspace{1cm}

\twocolumngrid

\bibliography{references}
\end{document}